\documentstyle[prl,twocolumn,aps,epsf]{revtex}
\begin{document}
\draft
\twocolumn[\hsize\textwidth\columnwidth\hsize\csname
@twocolumnfalse\endcsname

\title{Phase Separation Driven by External Fluctuations}
\author{J. Garc\'{\i}a--Ojalvo} 
\address{
Departament de F\'{\i}sica i Enginyeria Nuclear,
Universitat Polit\`ecnica de Catalunya,\\
Colom 11, E-08222 Terrassa, Spain}
\author{A.M. Lacasta}
\address{
Departament de F\'{\i}sica Aplicada,
Universitat Polit\`ecnica de Catalunya,\\
Av. Dr. Gregorio Mara\~n\'on, 50, E-08028 Barcelona, Spain}
\author{J.M. Sancho}
\address{
Institute for Nonlinear Science, Department 0407, University of
California, San Diego,\\
9500 Gilman Drive, La Jolla, California 92093-0407\\
and Departament d'Estructura i Constituents de la Mat\`eria,\\
Universitat de Barcelona, Av. Diagonal 647, E-08028 Barcelona, Spain}
\author{R. Toral}
\address{
Departament de F\'{\i}sica, Universitat de les Illes Balears and\\
Instituto Mediterr\'aneo de Estudios Avanzados (UIB-CSIC),
E-07071 Palma de Mallorca, Spain
}

%\date{\today}
\maketitle

%\begin{twocolumn}

\begin{abstract}
The influence of external fluctuations in phase separation processes
is analysed. These fluctuations arise from random variations of an
external control parameter. A linear stability analysis
of the homogeneous state shows that phase separation dynamics
can be induced by external noise. The spatial structure of the noise
is found to have a relevant role in this phenomenon. Numerical simulations
confirm these results. A comparison with order-disorder 
noise induced phase transitions is also made. 
\end{abstract}

\vskip2pc]

\vskip5mm
%\begin{twocolumns}
The role of noise as an ordering agent has been broadly studied in
recent years in the context of both temporal and spatiotemporal
dynamics. In the temporal case, which was the one to be addressed
earlier, external fluctuations were found to produce and control
transitions (known as {\em noise-induced transitions}) from monostable
to bistable stationary distributions
in a large variety of physical, chemical and
biological systems \cite{horsthemke84}. Spatiotemporal systems have
been faced much more recently. In these cases, the combined effects
of the spatial coupling and the noise terms acting upon the system
variables may
produce an ergodicity breaking of a bistable state, leading to phase
transitions between spatially homogeneous and inhomogeneous phases.
Results obtained in this field include critical-point shifts in
standard models of phase transitions \cite{broeck94a,ojalvo94,becker94},
pure {\em noise-induced phase transitions} \cite{broeck94b,broeck96,park96},
stabilization of propagating fronts \cite{armero96}, and noise-driven
structures in pattern-formation processes
\cite{ojalvo93,parrondo96,ojalvo96a,lythe96}.
In all these cases, the qualitative (and somewhat
counterintuitive) effect of noise is to enlarge the
domain of existence of the ordered phase in parameter space. It is the
purpose of this Letter to analyse the role of fluctuations in
a radically different type of spatiotemporal process, namely in
phase separation dynamics. It will be shown that external noise can induce the
phase separation process and that this effect is determined by the spatial
correlation of the noise terms. An important conclusion is that 
phase separation does not occur, necessarily, for the same
range of parameters for which the system presents a phase ordering process.

The dynamics of a large class of 
spatially-extended systems can be described in a
general way by the following standard model\cite{hh77}:
\begin{equation}
\label{general-model}
\frac{\partial \psi(\vec x, t)}{\partial t} = -{\cal L } (\nabla)  \left( \frac{
\delta {\cal F}}{\delta \psi} \right) + \xi(\vec x,t)
\end{equation}
where $\psi(\vec x, t)$ is a dynamical field that describes the state
of the system, $\cal F$ is a  
free energy functional, and $\xi(\vec x,t)$ is a
space-dependent stochastic process that accounts for thermal fluctuations.
Thermal equilibrium at temperature $T$ is reached at long times if the
fluctuation--dissipation relation holds:
\begin{equation}
\label{whitenoise}
\langle \xi(\vec x,t)\xi(\vec x',t') \rangle 
= 2 \varepsilon {\cal L} (\nabla) \delta(t-t') \delta(\vec x-\vec x')
\end{equation}
with $\varepsilon=k_BT$. The system is usually characterised by the
behavior of the spatial average of the field $\psi$, which plays the
role of an order parameter. The operator $\cal L (\nabla)$ does not alter the
equilibrium state of the system ($\sim \exp(-{\cal F}/k_BT)$), but only
its transient dynamics. Two forms of $\cal L$ are usually adopted:
for ${\cal L}=1$ (the so-called {\em model $A$}) the system evolves
towards its equilibrium state without any constraint on the value of
the order parameter; for $ {\cal L} (\nabla) = -\nabla^2$ ({\em model B}) the
order parameter is conserved throughout the dynamical evolution. Model 
$A$ is a prototype of ferromagnetic ordering, and model $B$ of phase
separation dynamics following a quench from a high temperature homogeneous
phase to a low temperature state. 
In this latter case, according to the value of the order
parameter and the quench location, the evolution might proceed either by
spinodal decomposition or by nucleation\cite{gunton}.

Far from the critical point, the dynamics is not qualitatively affected 
by the presence of the thermal noise term $\xi(\vec x,t)$.
Instead, we will consider the situation in
which there is an additional source of noise, $\eta(\vec x,t)$. 
This happens, for instance, when one of the
externally controlled system parameters is subjected to fluctuations.
As these fluctuations are external, they do not usually verify the
fluctuation--dissipation relation (\ref{whitenoise}) and the system is
no longer at equilibrium. We will show that,
in the case of model $B$, these external fluctuations can induce a phase
separation in the system. Although 
this is a reminiscence of noise--induced phase
transitions reported earlier for model $A$ \cite{broeck94a,becker94,ojalvo96b}
(and, in what follows, we will make a comparison between the effect of
external noise in both models),
an unexpected and notorious feature is that the nature of the destabilizing
terms due to the noise is intrinsically different for model $A$ and model $B$.
We conclude that
phase separation does not necessarily occur in the conserved
order--parameter model $B$ at the same values for which model $A$ shows
a noise--induced phase transition. It follows that, in contrast with what
happens at equilibrium, both models have now different stationary
distributions. Furthermore, a relevant 
result of our analysis is that the spatial structure
of the noise plays an important and distinct role
for both models. We assume the following correlation for the external noise
with a characteristic correlation length $\lambda$:
\begin{equation}
\label{colornoise}
\langle \eta(\vec x,t)\eta(\vec x\ ',t') \rangle 
= 2\sigma^2 \delta(t-t') g(|\vec x-\vec x\ '|/\lambda)
\end{equation}
where $g$ is a (short--ranged) spatial correlation function.
It is expected, and
will be confirmed in what follows, that since model $B$ represents
a domain-growth process, this correlation length will play a role far
beyond the intuitive intensity-reduction effect of space-time
noise correlation \cite{ojalvo94}. Correlation time of the noise
should not have a parallel influence, since all time scales of
the system are larger than those of the noise. Therefore, it seems
that a finite (non-zero) correlation time of the noise would not
differentiate between models $A$ and $B$, and will not be considered
here (the time dependence of the correlation of $\eta(\vec x,t)$ will be
assumed to be a Dirac delta, as shown in Eq.(\ref{colornoise})).

Although our results are quite general, we shall work, for the sake
of clarity, with the well known Ginzburg--Landau free energy
\begin{equation}
\label{free-energy}
{\cal F} = \int d\vec x \left[\frac{r}{2}\psi^2+\frac{1}{4}\psi^4 + 
\frac{K}{2} |\vec \nabla \psi|^2 \right]
\end{equation}
In the absence of additive (thermal) 
noise, phase separation occurs for $r<r_c=0$.
If additive noise is present, the transition occurs at $r_c < 0$. We
will consider $r>0$, so that order will not appear spontaneously. The
control parameter $r$ will be assumed to be subjected to external
fluctuations, {\em i.e.} $r\to r+\eta(\vec x,t)$. The spatial correlation
function $g$ is chosen to be a Gaussian of width $\lambda$:
\begin{equation}
\label{correlation}
g\left(\frac{|\vec{x}-\vec{x}\,'|}{\lambda}\right)=\frac{1}
{(\lambda\sqrt{2\pi})^d}\;
\exp\left(-\frac{|\vec{x}-\vec{x}\,'|^2}{2\lambda^2}\right)\, ,
\end{equation}
($d$ is the $\vec x$-space dimension)
which becomes a delta function in the limit $\lambda \to 0$.
It is simpler to analyze the role of noise by using a lattice discretization
in which the space vectors $\vec x$ take values $x_i$ ($i=1,\dots,N$), defined
on regular lattice of linear cell size $\Delta x =1$. The
field $\psi(\vec x,t)$ then becomes a discrete set of variables $\psi_i(t)$
and similar notation is used for the random fields  $\eta_i(t)$ and $\xi_i(t)$.
Under these considerations, the
lattice version of model (\ref{general-model}) with the Ginzburg--Landau
free energy (\ref{free-energy}) is:
\begin{equation}
\label{lattice-model}
\dot\psi_i=-{\cal L}_L(r\psi_i+\eta_i\psi_i+\psi^3_i-K\nabla^2_L
\psi_i)+\xi_i\,,
\end{equation}
where ${\cal L}_L=1$ for model $A$ and ${\cal L}_L=-\nabla_L^2$ for
model $B$. $\nabla^2_L$ is the lattice Laplacian operator.
Finally, in this version, the external noise has a correlation
function $g_{|i-j|}$ which is the discrete inverse Fourier transform of 
$\hat g_k$, the corresponding lattice version of the Fourier transform of 
(\ref{correlation}),
namely (in two dimensions):
\begin{equation}
\label{hatg}
\hat g_k = \exp\left(-\frac{\lambda^2}{2} \left(\sin(k_x/2)^2 +  
\sin(k_y/2)^2\right)\right)
\end{equation}
There is no closed analytical form for $g_{|i-j|}$ and the desired values
$g_0$ and $g_1$ (see later) must be obtained numerically.

The transition towards an ordered state can be analyzed by studying the
stability of the homogeneous phase $\psi_i=0$.
The early time evolution of the statistical moments of $\psi_i$ in Fourier space
can be obtained in a linear approximation. 
For example, 
the second moment (structure function) is defined as $S_k(t)=
\frac{1}{N}\,\langle \hat{\psi}_k \hat{\psi}_{-k} \rangle$, where
$N$ is the number of points of the system. 
Making use of the Stratonovich calculus and Novikov's theorem,
its evolution equation is \cite{ojalvo96a}: 
\begin{equation}
\frac{dS_k(t)}{dt} = -2\, \omega(k) S_k(t) + \frac{1}{N}\,f(k)\,
\sum_k \hat{g}_k S_k(t) +2\varepsilon
\label{structure}
\end{equation}
Hence the second moment equation contains a term which globally
couples Fourier modes and a constant term due to thermal
noise. The particular values of the dispersion relation $\omega(k)$
and of the mode-coupling coefficient $f(k)$ differ for models $A$
and $B$. For model $A$, the result is well known \cite{becker94}
\begin{equation}
\label{dispersion-A}
\omega^A(k)=r_{eff}^A + Kk^2,\;\;\;\;\;\;\;\;f^A(k)=1
\end{equation}
with an effective control parameter $r_{eff}^A=r-\sigma^2g_0$. For model $B$
the situation is drastically different:
\begin{equation}
\label{dispersion-B}
\omega^B(k)=r_{eff}^Bk^2+K_{eff}^B k^4,\;\;\;\;\;\;\;\;f^B(k)=k^2
\end{equation}
with effective control parameter 
$r_{eff}^B=r+\sigma^2 \nabla^2_Lg_0$ and effective diffusion coefficient
$K_{eff}^B=K-\sigma^2 g_1$. Two main
differences are observed with respect to model $A$: the diffusion
coefficient $K$ is also renormalized by the correlated external noise, 
and the noise-induced shift of the control parameter $r$ depends now,
through the Laplace operator $\nabla^2_L$, on the spatial structure
of the noise correlation, i.e. not only on the same--site correlation $g_0$,
but also on
the nearest--neighbor correlation $g_1$. 
These differences will reveal itself in the position of the
transition point where the homogeneous state loses stability and
phase separation appears.

When neglecting the mode--coupling terms in Eq.(\ref{structure}), it 
is readily seen that 
perturbations grow when $w(k)^{A,B}<0$ for some interval of $k$ values.
We have checked, by means of a numerical integration of Eq.
(\ref{structure}), that mode-coupling terms hardly influence the
position of the transition curves.
Hence the transition point is characterised by $r_{eff}=0$ for
both model $A$ and model $B$. The critical value of the control parameter
$r$ and its dependence on the spatial structure of the noise is,
however, different in the two cases:
\begin{eqnarray}
& &{\rm model ~}A:\;\;\;\;r_c=\sigma^2 g_0\nonumber\\
& &{\rm model ~}B:\;\;\;\;r_c=-\sigma^2 \nabla^2_Lg_0=\sigma^2 2 d(g_0-g_1)
\label{transition}
\end{eqnarray}
Figure 1 shows the transition curves between homogenous and inhomogeneous
states in the ($\lambda,\sigma^2$) plane for models $A$ and $B$
for a fixed value, $r=0.2$, of the control parameter and spatial dimension
$d=2$.
All points located above the curves shown in this phase diagram are
in an inhomogeneous state, which corresponds to an ordered phase in
model $A$ (solid curve) and to phase separation in model $B$ (dashed curve).
The $\lambda-$dependence of the model-$A$ curve is merely due to the
natural "softening" effect of noise correlation \cite{ojalvo94,ojalvo96a}.
In the case of model $B$, on the other hand,
additional, non-trivial dependence on the correlation length is
introduced via the Laplace operator. As a consequence, for small
$\lambda$, the transition in model $B$ occurs {\em sooner} in model
$B$ than in model $A$, whereas the situation is the opposite for large
values of $\lambda$. A crossing of the two transition curves occurs
for an intermediate value of $\lambda \approx 1.8$. We stress again that
the presence of ordered regions in the phase diagram of figure 1 is due
to the presence of a multiplicative noise on the model, since we are
taking $r=0.2$ which is larger than the mean--field critical value $r_c=0$.

The lines drawn in the phase diagram of Fig. 1 have been
obtained in a linear approximation (11). It is presumable
that this linear stability analysis will provide the position of the
transition points up to leading order of approximation \cite{becker94}.
In order to corroborate the results obtained by means
of the linear stability analysis, 
equations (\ref{lattice-model}) have been integrated numerically for
models $A$ and $B$ in dimension $d=2$. 
A standard stochastic algorithm
is used in order to handle both the additive and multiplicative noise
terms \cite{gard87}. Gaussianly-distributed random vectors are
generated by means of a numerical inversion method, optimised to
efficiently produce large quantities of Gaussian random numbers
\cite{toral93}. According to the previous discussion, the spatially 
correlated external noise is
generated in Fourier space with the desired correlation function (\ref{hatg}), 
and transformed back to real space at each integration time step
\cite{ojalvo96a}.

\begin{figure}[h]
\begin{center}
\def\epsfsize#1#2{0.40\textwidth}
\leavevmode
\epsffile{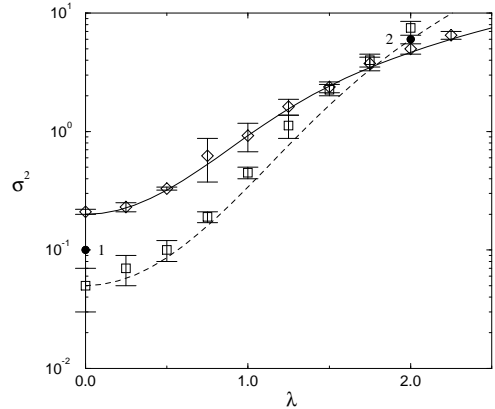}
\end{center}
\caption{
\label{fig1}
Phase diagram in the ($\lambda,\sigma^2$) plane for model A (solid
line) and model B (dashed line),
obtained from a linear stability analysis (8). Points correspond to
numerical simulations of the full model (6),
(diamonds: model A, squares: model B)
($r=0.2$, $K=1$, and $\varepsilon=10^{-4}$).
}
\end{figure}

A unified criterion for the existence of an ordered phase in model $A$
and of phase separation in model $B$ is the growth of the averaged second
moment of $\psi$ in real space (the averaged first moment is not
useful for model $B$ because this model conserves the order parameter). 
We define
this quantity as $J(t)=\frac{1}{N}\langle\sum_i\psi_i^2(t)\rangle$ or,
alternatively, as $J(t) = \sum_k S_k(t)$.
The instability point is thus defined so that, below it, $J(t)$ decays
to a thermal-noise background at large times and, above it, it grows
to a non-zero steady-state value $J_{st}$. In this way, one can
determine numerically the phase diagram of the system. The numerical
results are represented in Fig. 1 as diamonds (model $A$) and squares
(model $B$). It can be observed that the simulations of the full
nonlinear models reasonably adjust to the predictions of the linear
analysis. The agreement starts to fail at high values of $\lambda$
and $\sigma^2$. In fact, according to previous observations in
model $A$ \cite{ojalvo96b} and other models \cite{broeck94b,broeck96},
the transition curves might be expected to exhibit, for higher values
of the noise intensity, a reentrant branch towards the $\lambda=0$
axis. These reentrant transitions are hard to observe numerically
because of the large noise intensities involved.

Fig. 2 shows two patterns of a system evolving according to
model $B$, for point 1 
in the phase diagram of Fig. 1. Depending on the initial conditions we get
spinodal decomposition (Fig. 2a, with $\langle\psi(\vec x,0)\rangle=0 $), 
or nucleation (Fig. 2b, with $\langle \psi(\vec x,0)\rangle =0.1 $). 
For the same values of the noise
parameters a homogeneous phase is obtained for model $A$.

\begin{figure}
\begin{center}
\def\epsfsize#1#2{0.25\textwidth}
\leavevmode
\epsffile{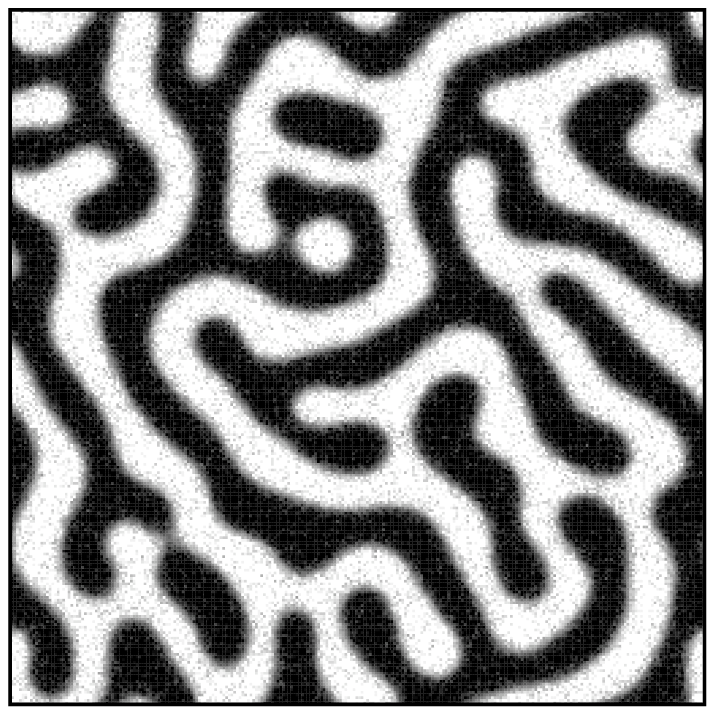}
%\end{center}
%\begin{center}
%\def\epsfsize#1#2{0.46\textwidth}
%\leavevmode
\hskip-5mm
\epsffile{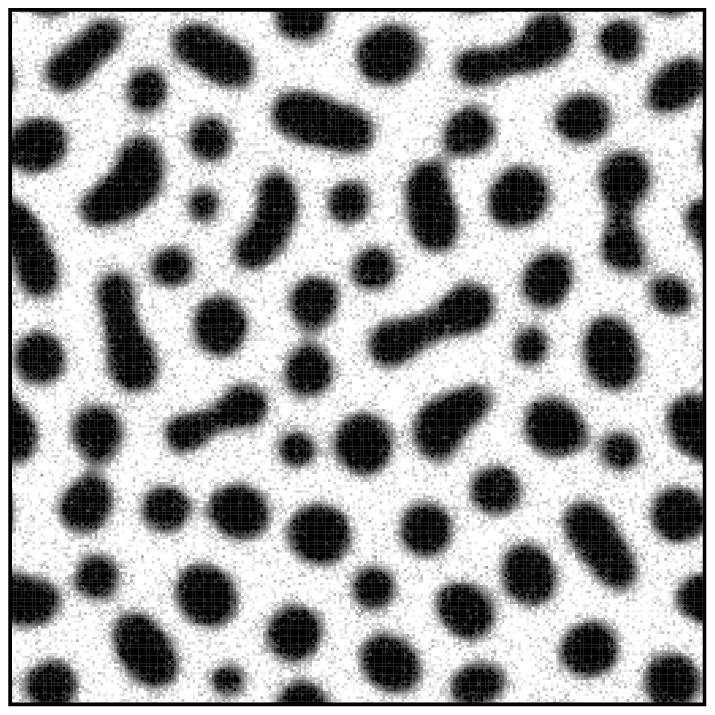}
\end{center}
\begin{center}
\vskip-5mm
(a)\hskip35mm (b)
\end{center}
\caption{
\label{fig2}
Spatial patterns of model B:  a) Spinodal decomposition and  b) Nucleation.
($t=2500$, $\lambda=0$, $\sigma^2=0.1$ and $\varepsilon=10^{-4}$)
}
\end{figure}

For larger values of the noise parameters a reverse situation is found.
Fig. 3 shows a spatial pattern of model $A$ for 
point 2 in the phase diagram of
Fig. 1. Now the homogeneous phase corresponds to model $B$.

\begin{figure}
\begin{center}
\def\epsfsize#1#2{0.25\textwidth}
\leavevmode
\epsffile{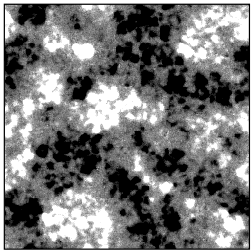}
\end{center}
\caption{
\label{fig3}
Spatial pattern of model A. ( $t=100$, $\lambda=2$, $\sigma^2=6$ and
$\varepsilon=10^{-4}$.)
}
\end{figure}

In conclusion, we have demonstrated for the first time the
ordering role of external noise in processes of phase separation.
The study has concentrated on the conserved time-dependent
Ginzburg-Landau model, although our results are not
restricted to this particular model. External noise is found to
enhance the phase separation process, and this effect is observed
to be modified by spatial correlation of the noise, which increases
the efficiency of fluctuations for small correlation lengths and
decreases it for large correlation lengths. Future work on this
issue should analyse the important 
effect of external noise on the dynamical
scaling of the phase separation process. Additional theoretical
analysis such as a mean-field approach \cite{broeck94a,broeck94b,broeck96}
could be useful in determining whether a reentrant transition should
be expected in this model.

We acknowledge financial support from the Direcci\'on General de
Investigaci\'on Cient\'{\i}fica y T\'ecnica (Spain), under projects
PB96-0421, PB94-1167 and PB94-1172), and computing support from
Fundaci\'o Catalana per a la Recerca.

%\end{twocolumns}


\begin{references}
\bibitem{horsthemke84} W. Horsthemke and R. Lefever, {\em Noise-Induced
Transitions} (Springer-Verlag, Berlin, 1984).
\bibitem{broeck94a} C. Van den Broeck, J.M.R. Parrondo, J. Armero, and
A. Hern\'andez-Machado, Phys. Rev. E {\bf 49}, 2639 (1994).
\bibitem{ojalvo94} J. Garc\'{\i}a-Ojalvo and J.M. Sancho, Phys. Rev. E
{\bf 49}, 2769 (1994).
\bibitem{becker94} A. Becker and L. Kramer, Phys. Rev. Lett. {\bf 73},
955 (1994); Physica D {\bf 90}, 408 (1995).
\bibitem{broeck94b} C. Van den Broeck, J.M.R. Parrondo, and R. Toral,
Phys. Rev. Lett. {\bf 73}, 3395 (1994).
\bibitem{broeck96} C. Van den Broeck, J.M.R. Parrondo, R. Toral, and
R. Kawai, Phys. Rev. E {\bf 55}, 4084 (1997).
\bibitem{park96} S.H. Park and S. Kim, Phys. Rev. E {\bf 53}, 3425
(1996).
\bibitem{armero96} J. Armero, J.M. Sancho, J. Casademunt, A.M. Lacasta,
L. Ram\'{\i}rez-Piscina, and F. Sagu\'es, Phys. Rev. Lett. {\bf 76},
3045 (1996).
\bibitem{ojalvo93} J. Garc\'{\i}a-Ojalvo, A. Hern\'andez-Machado, and
J.M. Sancho, Phys. Rev. Lett. {\bf 71}, 1542 (1993).
\bibitem{parrondo96} J.M.R. Parrondo, C. Van den Broeck, J. Buceta,
and F.J. de la Rubia, Physica A {\bf 224}, 153 (1996).
\bibitem{ojalvo96a} J. Garc\'{\i}a-Ojalvo and J.M. Sancho, Phys. Rev.
E {\bf 53}, 5680 (1996).
\bibitem{lythe96} G.D. Lythe, Phys, Rev. E {\bf 53}, R4271 (1996).
\bibitem{hh77} P.C. Hohenberg and B.I. Halperin, Rev. Mod. Phys. {\bf 49},435
(1977).
\bibitem{gunton} J.D. Gunton, M. San Miguel and P. Sahni, in {\sl Phase 
Transitions and Critical Phenomena}, vol. 8, edited by C. Domb and J.L.
Lebowitz, Academic, New York (1983).
\bibitem{ojalvo96b} J. Garc\'{\i}a-Ojalvo, J.M.R. Parrondo, J.M Sancho,
and C. Van den Broeck, Phys. Rev. E {\bf 54}, 6918 (1996).
\bibitem{gard87} T.C. Gard, {\em Introduction to Stochastic
Differential Equations} (Marcel Dekker, New York, 1987).
\bibitem{toral93} R. Toral and A. Chakrabarti, Comp. Phys. Commun.
{\bf 74}, 327 (1993).
\end{references}
\end{document}